\documentclass{emulateapj}
\usepackage{lscape,afterpage}
\usepackage{graphicx}

\shorttitle{GALEX M-star flares}
\shortauthors{WELSH et al.}

\begin{document}


\title{The detection of M-dwarf UV flare events in the $\it GALEX$ data archives}

\slugcomment{Accepted for the Astrophysical Journal Supplement {\it GALEX} Special Issue}

\author{
Barry Y. Welsh,\altaffilmark{1}
Jonathan M. Wheatley,\altaffilmark{1}
Mark Seibert,\altaffilmark{2}
Stanley E. Browne,\altaffilmark{1}
Andrew A. West,\altaffilmark{3}
Oswald H. W. Siegmund,\altaffilmark{1}
Tom A. Barlow,\altaffilmark{2}
Karl Forster,\altaffilmark{2}
Peter G. Friedman,\altaffilmark{2}
D. Christopher Martin,\altaffilmark{2}
Patrick Morrissey,\altaffilmark{2}
Todd Small,\altaffilmark{2}
Ted Wyder,\altaffilmark{2}
David Schiminovich,\altaffilmark{4}
Susan Neff,\altaffilmark{5}
R. Michael Rich,\altaffilmark{6}}

\altaffiltext{1}{Experimental Astrophysics Group, Space Sciences Laboratory, University of California, 7 Gauss Way, Berkeley, CA 94720-7450; bwelsh@ssl.berkeley.edu, wheat@ssl.berkeley.edu}

\altaffiltext{2}{California Institute of Technology, MC 405-47, 1200 East
California Boulevard, Pasadena, CA 91125}

\altaffiltext{3}{Dept. of Astronomy, University of California at Berkeley,
Berkeley, CA 94720-3411}

\altaffiltext{4}{Department of Astronomy, Columbia University, New York, NY 10027}

\altaffiltext{5}{Laboratory for Astronomy and Solar Physics, NASA Goddard
Space Flight Center, Greenbelt, MD 20771}

\altaffiltext{6}{Department of Physics and Astronomy, University of
California, Los Angeles, CA 90095}



\begin{abstract}
We present the preliminary results from implementing a new software tool that enables
inspection of time-tagged photon data for 
the astronomical sources contained within individual $\it GALEX$
ultraviolet (UV) images of the sky.  We have inspected the photon data contained
within 1802 $\it GALEX$ images to reveal
rapid,  short-term ($\la $ 500 sec) UV source variability in the form
of stellar `flares'. The mean
associated change in near UV (NUV) magnitude due to this
flaring activity is 2.7 $\pm 0.3$ mag.
A list of
49 new UV variable-star candidates is presented, together
with their associated Sloan Digital Sky Survey (SDSS) photometric magnitudes.
From these data we can associate the main source of these
UV flare events with magnetic activity on M-dwarf stars. Photometric parallaxes
have been determined for 32 of these sources, placing
them at distances ranging from approximately 25 to 1000pc. 
The average
UV flare energy for these flare events is 2.5 x 10$^{30}$ ergs, which
is of a similar energy to that of U-band, X-ray and EUV flares observed on many local 
M-dwarf stars.
We have found that stars of classes M0 to M5 flare with energies spanning a far larger range
and with an energy approximately 5 times greater than those of later (M6 to M8) spectral
type.

\end{abstract}




\section{Introduction}
The NASA Galactic Evolution Explorer ($\it GALEX$) satellite,
which was launched in April 2003, has been successfully obtaining imaging
photometric observations of astronomical
sources in two ultraviolet bands (near ultraviolet
[NUV] 1750 - 2750\AA, far ultraviolet [FUV] 1350 - 1750\AA). Scientific results from
the initial observation of galaxies and stellar sources can be found
in \cite{mar05} and references therein, and a description of the on-orbit instrumental
performance is described by \cite{mor05}. A major scientific data product
from $\it GALEX$ all-sky observations is a catalogue of sources detected in either/both
UV bands (see http://galex.stsci.edu/GR1/). These UV sources are
contained in the $\it GALEX$ Merged Catalog (MCAT) which can be accessed
at the Multi-mission Archive at the Space Telescope 
Science Institute (MAST). The catalogue is periodically updated and expanded as the
$\it GALEX$ mission progresses.
This catalogue contains a
wealth of information on both spectral and photometric
observations recorded by $\it GALEX$, but the most significant (and most widely used)
data-files are of the
combined-visit variety, which provide source positions and fluxes 
derived from the NUV and/or FUV channel observations.
A source visit refers to a
period of observation (i.e. an exposure) of a particular region
of the sky by $\it GALEX$, which can range from 
$\sim$ 100 seconds (for data recorded during the all-sky imaging survey)
to several tens of orbits (each lasting $\sim$ 1500 sec) recorded during the
deep imaging sky-survey mode of operation. 

Using the information contained in these source lists,
UV photometric variability $\it between$ observational visits to the same
position on the sky has been
found for some 84 objects, which are listed in the $\it GALEX$ UV
Variability (GUVV) catalog \citep{welsh05}. However, variability
in the form of small transient flux changes or short-lived flare events that may
possibly occur $\it within$ a single
observational visit by $\it GALEX$ can only be revealed by inspection of the
time-tagged photon data for individual sources. Since $\it GALEX$ has already detected 
several million UV sources during the first two years of its mission and
each detected photon event is time-tagged with a precision
of $<$ 0.05 seconds, clearly any search for
variability within all
these photon data files is extremely computer intensive and 
time-consuming. However, a new software tool
has recently been developed by the $\it GALEX$ Software Operations
and Data Analysis
team at Caltech that quickly enables source variability to
be revealed within each observational
visit (i.e. sky-field image) using the time-tagged photon count-rate data files. In this Paper we
report  the detection of 52 short-term UV variability events that
can be associated
with 49 stellar sources through
an inspection of 1802 individual $\it GALEX$ images.
All these 1.24$^{\circ}$ diameter sky-fields have Sloan Digital Sky Survey (SDSS)
DR4 imaging data \citep{adel06}
with associated multi-filter ($\it g$,$\it r$, $\it i$ and $\it z$) visible photometric magnitudes.
Using these
data we have been able to determine that the vast majority ($\sim$ 90$\%$) of these newly
detected $\it GALEX$ variable sources are M-dwarfs located up to 1kpc
from the Sun.

\section{Observations and Data Analysis}
Our initial search for variability within the time-tagged source photon lists was 
performed with data
generated from 1802 individual sky-field images contained
in the $\it GALEX$ data release IR1.1+GR1, which is
soon to be made publicly
available at the MAST facility. 
We have restricted our present study to data 
recorded in
the NUV band through pointings  made during (a) the
$\it GALEX$ Medium Imaging Survey (MIS), (b) the
Deep Imaging Survey (DIS) and (c) the Nearby Galaxies Survey (NGS)
 \citep{mar05}. Only image fields that had associated SDSS  
visible imaging photometric data were chosen for this study. 
Data from the $\it GALEX$ All-Sky Imaging
Survey (AIS) were not included due to the  short duration of
these exposures.
Although similar photon data  exists in the FUV wavelength
band for the MIS, DIS and NGS observations, those data
are of an inferior S/N ratio to the NUV channel and are not presented here.

Each of
the 1802 $\it GALEX$ images has a diameter of  $\sim$ 1.24$^{\circ}$ on
the sky, and each exposure was of $\sim$ 1500 seconds duration (i.e.the
length of one $\it GALEX$ orbital eclipse). For sky-fields observed in the MIS and NGS modes
usually $\la$ 2 exposures per sky-field were recorded, but for several fields observed in
the DIS mode as many as 50 exposures were recorded at a particular
position on the sky. In a few cases, the exposures were shortened to
$\sim$ 200 seconds due to on-orbit operational constraints. The data set is not
uniform in the sense that many exposures of the same sky-field were taken
several months apart. In addition, all consecutive exposures of the same
field have a 60 minute gap between observations due to the $\it GALEX$ satellite
orbital eclipse period.
All of these data correspond to observations of 752 different
regions of the sky, with a total on-sky integration time
of 2.27 x 10$^{6}$ seconds. In Figure 1 we show the Galactic distribution of
the 1802 $\it GALEX$ MIS, DIS and NGS
image fields that have been used for our
present data analysis. These image fields were primarily selected
for the study of external galaxies and hence the vast
majority of the fields are located at moderately
high Galactic latitudes, well away from the Galactic plane. 

The raw UV imaging data were 
processed using the standard $\it GALEX$ Data Analysis Pipeline (version 4.0) operated
at the Caltech Science Operations Center (Pasadena, CA), which inputs time-tagged
photon lists, instrument and spacecraft housekeeping data and
satellite pointing aspect information \citep{mor05}. The
data pipeline then uses the automated SExtractor source detection
algorithm \citep{bertin96} to produce a final catalog of 
source positions on the sky with corresponding ultraviolet magnitudes (averaged
over the entire duration of the exposure)  for
each observation. At present, the time-tagged photon lists associated with
each of these sources have not been widely available to the astronomical community.
However a new software tool (named `varpix'),
now allows post-pipeline inspection of the individual photon files for each source
in order to search
for any time-variability present during each $\it GALEX$ exposure.

Since each of the 1802
exposures contains $\sim$ 1 Gbyte of data, clearly some sort of
data-compression routine is required to search for source
variability within the terra-byte of available photon data. 
`Varpix', re-bins
each of the normal 1.5 arcsec$^{2}$ $\it GALEX$ image pixels into
12 arcsec$^{2}$ pixels and sums the photon counts in each of these larger sky-pixel areas
over consecutive 16 second intervals.
For every large sky-pixel (i) the median value
of the summed photon counts, medpix (i,j) is calculated for
each frame (j) such that a median image is created for
all the frames. Similarly, for every sky-pixel (i) a maximum photon count,
maxpix(i,j) is calculated for each frame, thus creating a maximum image of
all frames. The variation, varpix(i), of the counts in
sky-pixel (i)  compared with the median and maximum image values
is computed as:

varpix(i) =  [maxpix(i,j) - medpix(i,j)] / sqrt(medpix(i,j))

This algorithm is
essentially a crude photon count `variability signal-to-noise' estimator 
that is calculated for each (large) sky-pixel image.
The final product is thus a map of
sky-pixel count-variability determined over a period of (j) 16-second long frames.
The source detection algorithm, SExtractor, was then run on
any of the large sky-pixel images with values of
varpix(i) $\gtrsim$ 15 to produce an initial list of possible
variable source candidates. These positions were then matched to
the spatially nearest astronomical source listed in the standard merged
catalog (MCAT) of $\it GALEX$ UV sources to verify their reality.
In several instances,
other orbiting satellites and asteroids were revealed by this screening process, and
are not reported in our present list of variable stellar sources.
The varpix(i) $\gtrsim$ 15 value is an empirical signal-to-noise
constraint that, when coupled to restricting the sampling of
an image to the central 0.5$^{\circ}$ radius of
the detector, eliminates spurious
detections caused by instrumental edge reflections and glints that
can contaminate an image. 
Extremely bright UV stellar sources,
whose halo-broadened point spread function
(which during $\it GALEX$ observations is dithered in a spiral pattern
on the sky)
can sometimes extend beyond the 12 arc-second
sized sky-pixel area and are also filtered out.

The time-tagged photon data associated with each of the
the confirmed variable stellar sources were then extracted from
the large sky-pixel images contained within each of
the 16 second-long varpix frames, and (after background subtraction)
were adaptively binned to produce time intervals in
which the source photon flux reached a S/N
ratio = 10. The source photon data contained in these bins were
then converted into values of NUV magnitude, m$_{nuv}$,
in the AB magnitude system of
\cite{oke90} using the $\it GALEX$ photon flux conversion of
\cite{mor05}. The resulting data product was set of values
of m$_{nuv}$ as a function of adaptively binned time (i.e. a light-curve) for each
variable source. In Figure 2
we show examples of these light-curves for nine of the
newly detected variable sources. The physical nature
of these short-term variations in flux will be discussed in the following
Section.  

The present `varpix'  algorithm is capable
of detecting stellar flares
that exhibit changes in the light-curve
values of NUV magnitude ($\Delta$m$_{nuv}$) of $>$ 0.25 mag 
on sources with a quiescent magnitude as faint as m$_{nuv}$ $\sim$ 21.0.
We note that the detection sensitivity for variable
sources is $\it not$
equivalent to that of the $\it GALEX$ (stable) source detection limit of
m$_{nuv}$ = 22.7  listed
by \cite{mor05}. The inferior sensitivity for variable source detection using
`varpix'
is mainly due to the division of source photons into many 16-second bins, as
opposed to the normal $\it GALEX$ stable source sensitivity
which is derived from a single 1500 second accumulation of photons during
an MIS observation.
Although the present `varpix' detection algorithm is
biased against the inclusion of very bright flare events, in fact
none were found in these data (as subsequently confirmed by a manual search through all the
exposures).
Thus,
only sources with with derived values
of m$_{nuv}$
that lie in
the magnitude range 14.7 - 21.0 were presently confirmed as being variable. 

We also note that variable sources detected
in the  NUV channel were all subsequently confirmed
as being variable in the FUV channel data, often with a comparable
variation in the magnitude of their FUV flux. 
Although the variability detection process is 
incomplete for faint sources and
low levels of variability, it presently provides a computationally fast
and easy way of revealing astronomical sources that vary significantly
over short time-periods ($<$ 250 seconds) at UV wavelengths. Work is  
in progress to further refine all of the various selection criteria
and screening constraints
in order to produce an more complete catalog of UV variable sources.

In Table 1 we list information on
the 49 variable UV sources (associated with
52 separate flare events) that have initially been found in the
1802 image fields using the software detection tool described above.
Two separate UV transient variability events were
recorded in different images for 3 of the sources, SDSS J100141.6+020758.7,
SDSS J145110.28+310639 and SDSS J171746.57+594124.1.
In columns (1) and (2)
of this Table we list 
the Right Ascension (J2000.0) and Declination (J2000.0) of
each detected source
as measured by $\it GALEX$.
Typically, these positions are accurate to $\pm$ 1.5 arc sec for detections in the central
1$^{\circ}$ of the $\it GALEX$ detectors \citep{mor05}. In column (3) we list the
nearest stellar source to this position, as catalogued in the
SDSS  DR4 catalog \citep{adel06}. For very close binary systems the
brighter companion was always chosen, and in
one case
the association with an SDSS  identifier number was not possible due to the
SDSS  image being contaminated by a cosmic ray streak.
In column (4) we list the
MAST identifier for the respective
$\it GALEX$ image field in which the variability event was discovered.
In column (5) we list the number of observations, N$_{obs}$, for each source
contained within the $\it GALEX$ archive that were
searched using the `varpix' software tool.
In columns (6) and (7) we
tabulate the maximum NUV magnitude (m$_{NUVmax}$) observed
within the duration of the flare and the
associated change in this
magnitude ($\Delta$m$_{NUV}$) from the faintest value measured during the length
of the observational visit. For the
majority of sources this faint magnitude value is not equal to
the true quiescent NUV magnitude, since
for $\sim$ 60$\%$ of the sources the lowest value of m$_{nuv}$
recorded in most of their exposures lay below the detection threshold
of the NUV channel (i.e. m$_{nuv}$ $>$ 22.7).
Columns (8), (9), (10)
and (11) list the SDSS  visible point-spread function
magnitudes in the $\it g$, $\it r$,
$\it i$ and $\it z$ photometric bands \citep{abaz03}. 
Magnitude values marked with asterisks are subject to
saturation effects and have been excluded from any
subsequent calculations and conclusions in this Paper.
Finally, in column (12) we provide a probable spectral
type for each 
source based on its SDSS  photometric colors,
or (for a few very bright objects) a catalog name based on cross identification
with the Simbad on-line catalogue of sources.
We have mainly
used the ($\it r$ - $\it i$) colors of
\cite{west05}, supplemented
where necessary by the ($\it g$ - $\it r$) colors of
\cite{aguer05} and the ($\it i$ - $\it z$) colors
of \cite{west05}, to assign the spectral types.

\section{Discussion}
\subsection{The Light-Curves}
A detailed inspection of all 52 light-curves, of which Figure 2 shows nine
typical examples, reveals that the short-term variability events can broadly be described by three
different signatures, which are very similar to those revealed from numerous ground-based
U-band observations of M dwarf flares (\cite{moffett76}, \cite{houd03}).
Type 1 events,
shown in Figures 2(a) - (c), consist of a rapid flux rise (typically $<$ 50 seconds)
with a single emission peak that is followed by an 'quasi-exponential' decay (typically
lasting $<$ 500 seconds) that normally returns to
the pre-event flux level. Type 2 events, shown in Figures 2 (d) - (f), consist
of a similarly rapid flux rise, but have multiple secondary emission peaks that eventually decay to
the pre-event flux level after $>$ 500 seconds. Type 3 events,
shown in Figures (g) - (i), have
more complex rising and falling flux signatures. Of the 52 variability curves
that $\it GALEX$ has observed, 
27 can be classed as Type 1 events, 15 are of Type 2,
4 are of Type 3 and the remaining 6 events cannot be classified
due to an incomplete time sampling of their
time versus flux-signatures (i.e. they occur too close to the start or end of an exposure).
The similarity between the general shapes of the NUV and U-band flare light-curves suggests
that a common emission mechanism is responsible for the observed flux. It
has been well-established that U-band and bluer wavelength
flare spectra are dominated by continuum
emission \citep{hawley03}, which strongly favors
a similarly significant continuum contribution to the flare spectra
recorded at NUV wavelengths by $\it GALEX$. The relative
contribution from continuum and/or emission lines in the
$\it GALEX$ UV bands recorded during a flare
has been discussed by \cite{robinson05}, who also argue
that the NUV band is dominated by stellar continuum flux. However,
we note that for the $\it GALEX$ FUV band, line emission also becomes
an important contributor \citep{phillips92}.

In Figure 3 we show a histogram plot of the change in
NUV magnitude measured over one
 $\it GALEX$ exposure, $\Delta$m$_{nuv}$, versus the number of 
detected variable sources, N, that exhibit that magnitude change.
The distribution
peaks close to the
mean value of $\Delta$m$_{nuv}$ = 2.7$\pm 0.3$ magnitudes (i.e. a change of greater
than a factor 10 in the emitted UV flux). This plot takes no account
of the spectral type or distance to each flare star
and is presented as an empirical example of the range of changes
in NUV magnitude that the $\it GALEX$ satellite can typically encounter
during its scheduled observations. 
We note the presence of  
a subsidiary second peak in this distribution (albeit
with a small number of events), that spans a 4.5 - 5.5 magnitude change.
Events of this nature are characterized by an increase in flux of $>$ 100,
and seem much rarer than the former less energetic events. For comparison purposes
we note that the very large UV flare on GJ 3685A detected with $\it GALEX$
had a value of $\Delta$m$_{nuv}$ = 7.7 magnitudes and
a derived energy of $\sim$ 10$^{34}$ ergs  \citep{robinson05}. 

\subsection{M-dwarf flare energies and flare frequency}
In Figure 4 we plot values of  ($\it r$ - $\it i$) versus ($\it i$ - $\it z$) for the sources
listed in Table 1 that possess reliable (i.e. unsaturated) SDSS photometric magnitudes.
It is clear from this plot that the great majority of variable sources
have values of ($\it r$ - $\it i$) $>$
0.8 and ($\it i$ - $\it z$) $>$ 0.4. All these sources can be categorized
as M-dwarf stars, based on the colors for cool stars
presented in  \cite{fin00} and \cite {west05}. The remaining UV variable sources
are most probably active K-dwarfs. Several of the variable objects appear as binary systems
in the SDSS  images, with the source SDSS J101152.3+614454.1 being a triple
M-star system.  
The data presented in
Figures 2 and 3 also support the notion that the vast majority of the changes in
UV flux originate in a single physical type of astronomical source, i.e. stellar flare
eruptions on K and M stars. Since M-dwarfs account for more than 75$\%$ of the stellar
population in the solar neighborhood, and such stars  
are known to 
possess strong magnetic fields with high coronal activity
and associated chromospheric UV line emission \citep{mitra05}, it is therefore
not surprising that $\it GALEX$ observations
favor their serendipitous detection.

Unfortunately the S/N ratio of the data for the majority of these short-term flare events
precludes us from performing a detailed physical analysis similar to that carried
out on the giant UV flare observed by $\it GALEX$ on the star
GJ 3685A \citep{robinson05}. However, since we have a relatively large
sample of M-dwarf flares we shall proceed with a
more general statistical treatment of the present data.

In order to estimate the total emitted UV energy from each of each of these flares we
require both knowledge of the integrated flux emitted over the time of the flare event,
as well as the distance and stellar bolometric luminosity for each source. 
Distances to all of the M-dwarfs that have
good quality SDSS photometry were derived using the photometric parallax
method described in \cite{west05}. We have determined
that 32 of the 49 sources listed in Table 1 are both M-dwarfs $\it and$ have useable $\it i$
and $\it z$ band data (we preclude the J101152.3+614454.1 triple
star system). Absolute magnitudes have been calculated for these 32 stars
from the $\it i-z$ color relationship, with photometric distance
estimates being subsequently derived.
The corresponding stellar bolometric luminosity for each of these M-dwarfs is based
on their spectral type (as listed in Table 1), and is estimated from the
luminosity values given for individual M-dwarfs of similar spectral-type
presented in \cite{legg96,legg01}. 
Due to our uncertainty of $\sim$ one spectral
class in the values listed for each source in Table 1, the derived
value of bolometric luminosity for each star, L$_{bol}$, is an average
of the two luminosities for both of the spectral classes listed.
In Table 2 we list our derived values of distance and L$_{bol}$ (compared to that of
the solar luminosity, L$_{\sun}$). In addition we list the total NUV flare energy,
E$_{NUVflare}$  and the NUV flare luminosity, L$_{NUVflare}$, 
referenced to the stellar
bolometric luminosity, L$_{bol}$. Note that two flares were detected on the
star J100141.6+020758.7, such that flare energies for 33 separate events are listed
in Table 2. 

The present sample of stars whose
distances can be derived from photometric colors is biased against the inclusion of bright
and nearby M dwarfs, due to a bright limit on SDSS magnitudes. However,
for the 32 stars who do have distance estimates, flares were
detected over
the 25 - 1000pc range. The majority of these stars are located
$<$ 300pc from the Sun, thus placing them within the thin
disk. We note that stars
of spectral type M5/M6 and later were only detected with distances
$\la$ 200pc. This is most probably due to an apparent magnitude selection
effect, in which the intrinsically brighter and earlier spectral types
(M0 - M4) were detectable to greater distances with $\it GALEX$. 
The most distant flare recorded was on the star J023955.52-072855.4 ($\it l$ = 181$^{\circ}$,
$\it b$ = -58$^{\circ}$ ) with a
distance of 990pc, thus placing it $\sim$ 840pc below the Galactic plane. 

The range of the 33 UV flare energies detected by $\it GALEX$ extends from
1.2 x 10$^{28}$ to 1.6 x 10$^{31}$ ergs, with an
average UV flare energy of 2.5 x 10$^{30}$ ergs. This energy range is very similar
to that derived from NUV observations of 54 flares
on the dM4.5e star YZ Canis Majoris by \cite{robinson99}.  
We also note that the presently derived NUV flare energies are
similar to those of the U-band flares observed at visible wavelengths
on UV Ceti \citep{panagi95}, but they do not reach the more
extreme flare energies recorded in the U-band by \cite{petter84} for
the nearby dM3.5V star AD Leonis. Our derived UV flare energies
are also similar to those reported for M-dwarf flares detected
at EUV and X-ray wavelengths by \cite{sanz02}, \cite{gudel04} and \cite{mitra05}.

The superposition of flares with a large
range of energies
is thought to play a fundamental role in the coronal (X-ray) heating of magnetically
active dMe stars \citep{gudel04}, and thus studies of the flare frequency
as a function of flare energy can be important in determining
if flares are sufficiently numerous and energetic to
explain the associated coronal emission \citep{audard00, gudel03}.
Table 2 shows that
there is a distinct difference in the range of UV flare energies for different spectral-types of
M-dwarf. For example, flares detected on the 28 stars with
spectral classes M0 to M5 were
detected with energy ratios spanning the range log (L$_{NUVflare}$/L$_{bol}$) = -2.4 to -5.5,
with a median energy of log (L$_{NUVflare}$/L$_{bol}$) = -3.6. However, for the (albeit
small) sample of 5
stars of spectral classes M6 to M8 the mean flare energy spans a much smaller
energy range of log (L$_{NUVflare}$/L$_{bol}$) = -4.2 to -5.7, with a median flare energy
of log (L$_{NUVflare}$/L$_{bol}$) = -4.3 (i.e. $\sim$ a factor 5 lower in energy).
This pattern of behavior is similar
to that reported by \cite{west04} and \cite{cruz02}, in which the activity strength (as
measured by the ratio of H$\alpha$ luminosity to the stellar bolometric
luminosity) was also found to be higher for stars of spectral class
M0 to M5 than for those of type M6 to M9. We refer
the reader to \cite{gizis02} for a discussion of
why the level of chromospheric activity may vary with M-dwarf stellar
age and mass.

A derivation of the (UV) flare frequency rate on M-dwarfs is also particularly important for the
study of habitability zones on possible associated extra-solar planetary systems
\citep{turnbull03}. Previous estimates of flare
frequency have been made by \cite{smith05} who observed 17 large X-ray
flares during $\sim$ 120,000 seconds observations of 5 dMe stars
using the $\it XMM$ satellite (i.e. 0.4 X-ray flares hr$^{-1}$ M-star$^{-1}$). Also, 
\cite{mitra05} have observed 13 UV (and near-simultaneous X-ray) flares during 40
hours of observations of 5 dMe stars
using the Optical Monitor (used with NUV filters) also on the $\it XMM$ satellite (i.e.
3.08 UV/X-ray flares hr$^{-1}$ M-dwarf$^{-1}$). In addition, 
 \cite{gudel04} observed
almost continuous low-level X-ray variability during
65 ksec of $\it XMM$ observations of Proxima Centauri (dM5.5e). 
Unfortunately, although our $\it GALEX$ observations
have sampled a relatively large number of sky-fields for a relatively
extended period of time (from which, in principle, a UV flare-rate per
M-dwarf could be derived), the majority of the sight-lines are located 
well above the Galactic plane where                      
the M-dwarf space density is poorly determined.
As the $\it GALEX$ mission progresses with an increased sky-coverage
for lower Galactic latitudes,
we will then be better placed to derive a statistically meaningful M-dwarf flare rate.
In particular, $\it GALEX$ extended observations of the entire Hyades star cluster (that
covers $\sim$ 30 sq.deg on the sky) may provide the best estimate of flare frequency
for local M-dwarfs.

\subsection{Conclusion}
We describe a newly developed software tool that allows inspection of the
time-tagged photon lists associated with sources
contained within each $\it GALEX$ $\sim$ 1.24$^{\circ}$
diameter UV image of the sky. In its present form, potential UV source variability can
only be reliably detected by this algorithm for objects with
NUV magnitudes in the range 14.7 $<$ m$_{nuv}$ $<$ 21.0.
A preliminary inspection of 1802 $\it GALEX$ NUV
images, recorded over a total exposure time of 2.27 x 10$^{6}$ seconds, has revealed
52 short-term transient UV outbursts originating on 49 different
stellar sources. A comparison of the SDSS  $\it g$, $\it r$,
$\it i$ and $\it z$ visual photometric magnitudes for these stellar sources
shows that the vast majority can be categorized as nearby active M-dwarfs.
Such stars are known to possess strong magnetic fields with high
coronal activity and thus we can confidently associate the transient UV outbursts
with stellar flares.

The light-curves for these 52 variability events can be broadly described
by 3 different flux-signatures, with $\sim$ 50$\%$ of the outbursts
consisting of a rapid flux rise (typically $<$ 50 seconds) with a single
emission peak that is followed by an `exponential-like' decay that typically
lasts $<$ 500 seconds. This behavior is similar to many flares observed in the U-band on
numerous M-dwarfs, whose spectra are dominated by continuum emission.
It therefore seems highly likely that the flare emission
recorded in the $\it GALEX$ NUV band is similarly dominated by continuum emission. 

The mean NUV magnitude change for
these short-lived flare events
is 2.7 $\pm 0.3$ mag., with several outbursts being $>$ 10 times more
energetic than the mean value. Photometric distances in the 25 - 1000pc
have been derived for flares observed on 32 of these M-dwarfs. 
The corresponding average
NUV flare energy for the flare events on these
stars of known distance is 2.5 x 10$^{30}$ ergs, which
is of a similar energy to that of U-band, X-ray and EUV flares observed on many local dMe stars.
We have found that stars of classes M0 to M5 flare with energies spanning a far larger range
and with an energy $\sim$ 5 times greater than those of later spectral type (i.e. M6 to M8).

Finally, we note that although the utility of the new `varpix' software tool has been
demonstrated on $\it GALEX$ UV photon data associated with M-dwarf flare events, 
it could also be used to reveal lower level short-term variations in the UV flux from
other astronomical objects such as cataclysmic variable stars, BL Lac and Seyfert galaxies.

\begin{acknowledgements}
$\it GALEX$ (Galaxy Evolution Explorer) is a NASA Small Explorer, launched in April 2003.
We gratefully acknowledge NASA's support for construction, operation,
and science analysis for the GALEX mission,
developed in cooperation with the Centre National d'Etudes Spatiales
of France and the Korean Ministry of
Science and Technology. We acknowledge the dedicated
team of engineers, technicians, and administrative staff from JPL/Caltech,
Orbital Sciences Corporation, University
of California, Berkeley, Laboratoire d'Astrophysique de Marseille,
and the other institutions who made this mission possible. 
Financial support for this research was provided by
the NASA $\it GALEX$ Guest Investigator science program.
This publication makes use of data products from the SIMBAD database,
operated at CDS, Strasbourg, France.
\end{acknowledgements}


\begin{figure}
\center
{\includegraphics[height=10cm]{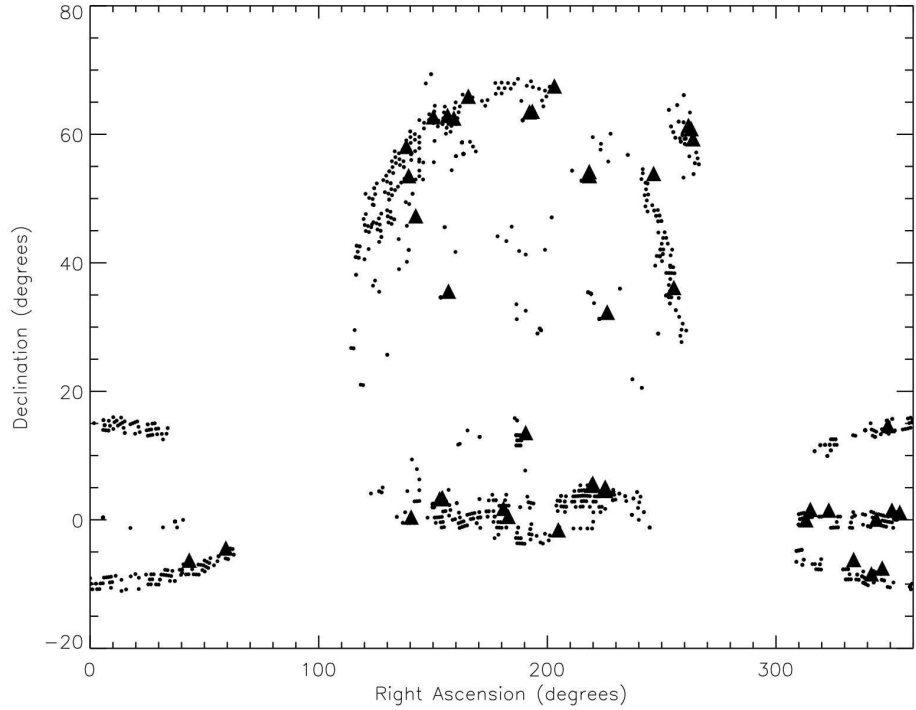}}
\caption{Sky distribution of the $\it GALEX$ image fields (filled circles) used in this analysis. Filled triangles = fields that exhibited UV flare activity. } 
\label{Figure 1}
\end{figure}

\begin{figure}
\center
{\includegraphics[height=16cm]{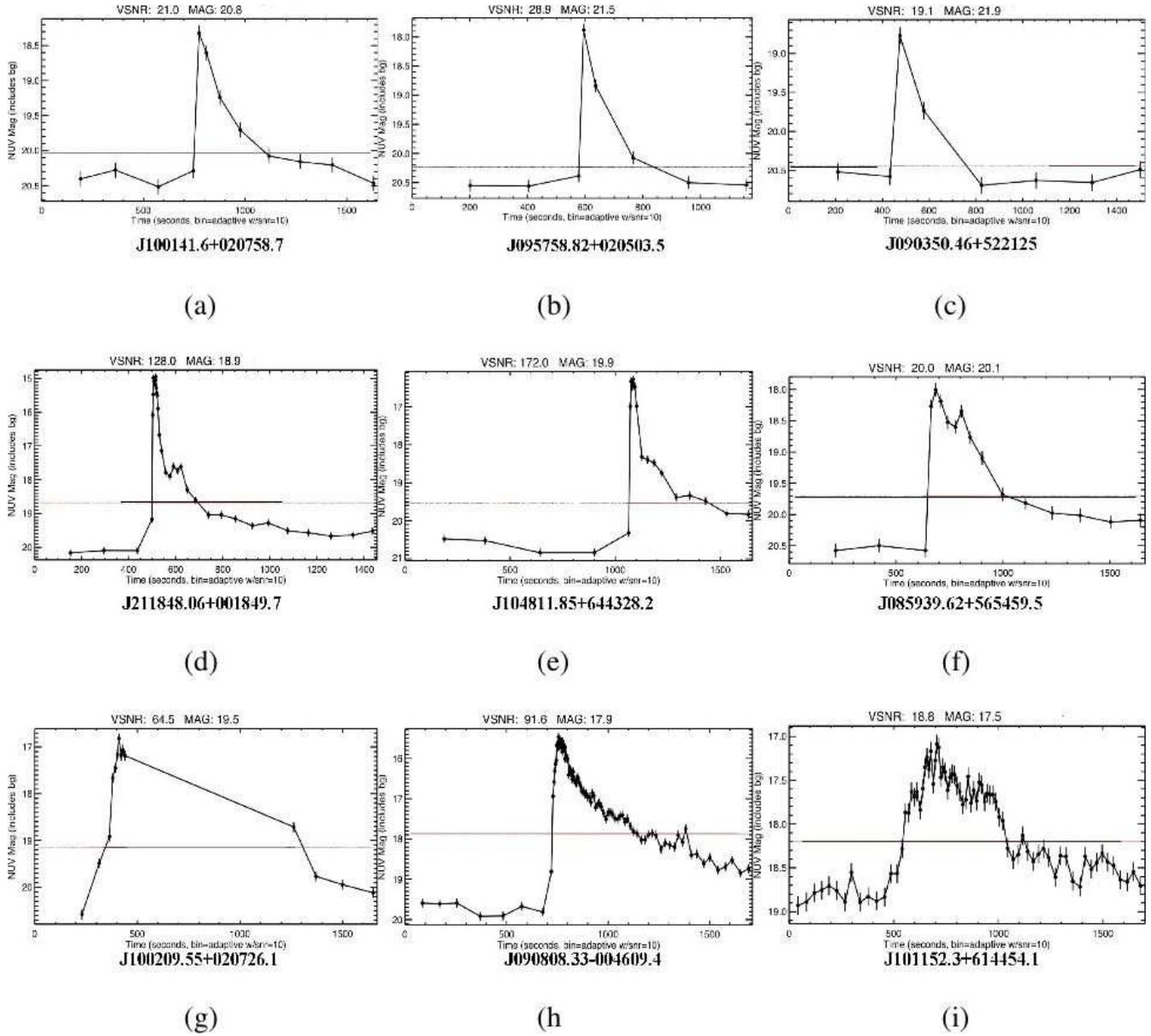}}
\caption{UV light curves of the observed transient increase in NUV magnitude, m$_{nuv}$, versus time for several sources. Points are plotted at adaptively-binned time intervals with an associated UV flux of S/N ratio =10 (see text for details). The listed VSNR value on each plot is the peak S/N ratio, varpix(i), of the transient event at the peak NUV magnitude. The value of MAG refers to the NUV magnitude listed in the $\it GALEX$ merged catalog of sources (MCAT) and represents the mean magnitude averaged over that whole exposure. The faint line across each plot is the median value of UV flux recorded over the length of the exposure. The various light-curve shapes are discussed in the text.} 
\label{Figure 2}
\end{figure}

\begin{figure}
\center
{\includegraphics[height=10cm]{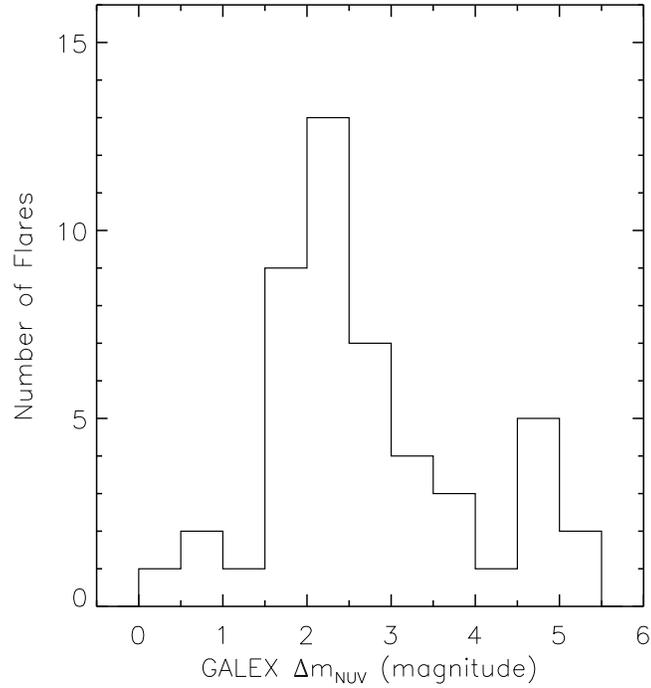}}
\caption{Histogram distribution of the number of UV flares with an associated change in NUV magnitude of $\Delta$m$_{nuv}$ observed during one $\it GALEX$ exposure. No account has been made to separate the effects of the different spectral types of M-dwarf flare.} 
\label{Figure 3}
\end{figure}

\begin{figure}
\center
{\includegraphics[height=10cm]{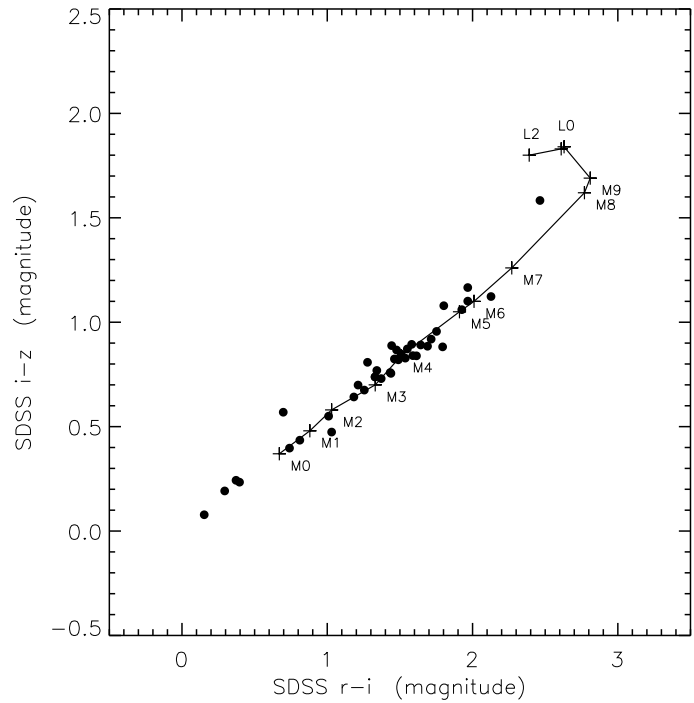}}
\caption{Plot of SDSS ($\it i$ - $\it z$) versus ($\it r$ - $\it i$) photometric magnitudes for the UV variable sources (filled circles) listed in Table 1. Superposed is the color-color relationship (line joining the `+' points) derived by West, Walkowicz and Hawley (2005) for spectral types M0 to L2. The four stars to the blue of the M dwarf locus lie in the color-color region of late K stars.} 
\label{Figure 4}
\end{figure}

\afterpage{
\begin{landscape}
{
\renewcommand{\arraystretch}{0.7}
\begin{table*}[htbp]
\caption{\label{obslog} List of short-term UV variable objects }
\begin{tiny}
\begin{flushleft}
\begin{tabular}{llcccccccccc}
\hline
\hline
R.A. (2000.)&Decl. (2000.)&SDSS I.D.&$\it GALEX$ Field & N$_{obs}$&m$_{NUVmax}$&$\Delta$m$_{nuv}$&$\it g$&$\it r$&$\it i$&$\it z$&Spactral Type (or I.D.)\\
\hline
39.981728&-7.481963&J023955.52-072855.4&MISDR1-18474-0455-001&5&18.66&2.08&19.13&17.76&16.95&16.52& M0/M1 \\
50.021702&-7.043504&*cr&MISDR1-28516-0459-0001&2&17.57&2.69&N/A&N/A&N/A&N/A&N/A\\
55.991768&-5.590886&J034358.01-053526.5&MISDR1-26995-0462-0004&6&19.57&0.94&21.01&19.47&17.89&17.00&M3/M4 \\
134.915399&56.916793&J085939.62+565459.5&MISDR1-03050-0448-0001&2&18.00&2.58&16.70&15.28&14.25&13.78&M1/M2\\
135.959594&52.356791&J090350.46+522125&MISDR1-03245-0552-0001&1&18.77&1.92&23.67&21.75&19.63&18.50&M5/M6\\
137.034690&-0.769202&J090808.33-004609.4&MISDR1-24351-0470-0002&4&15.47&4.45&15.40&14.52&14.13&13.89&late K\\
139.034229&46.123054&J091608.29+460724.5&MISDR2-03559-0832-0001&2&18.80&1.86&17.31&15.78&14.09**&13.20&M4/M5\\
146.746022&61.529637&J094659.05+613146.2&MISDR1-00464-0486-0001&4&17.20&3.86&22.57&21.10&19.14&17.97&M5/M6\\
149.495282&2.084214&J095758.82+020503.5&COSMOS-04-0022&120&17.88&2.68&19.09&17.62&16.03&15.19&M3/M4 \\
150.423269&2.133063&J100141.6+020758.7&COSMOS-00-0003&169&17.26&3.17&17.60&16.14&14.81&14.07&M2/M3\\
150.423667&2.133001&J100141.6+020758.7&COSMOS-02-0053&169&18.32&2.19&17.60&16.14&14.81&14.07&M2/M3\\
150.532760&2.376672&J100207.86+022234.7&COSMOS-02-0034&116&16.84&3.38&17.06&15.60&15.32**&13.56&early M\\
150.539877&2.124081&J100209.55+020726.1&COSMOS-01-0035&114&16.82&3.76&21.51&19.99&18.50&17.68&M3/M4\\
150.731214&2.199898&J100255.43+021159.6&COSMOS-02-0023&107&19.37&1.25&20.93&19.52&18.33&17.69&M2/M3\\
152.967014&61.749734&J101152.3+614454.1&MISDR2-00492-0771-0001&7&17.09&1.84&22.17&19.82&18.40&17.17&M3/M4 triple star system\\
153.361965&34.386136&J101326.83+342309.8&UVE-A0951-0004&38&18.14&2.71&22.34&20.74&19.23&18.38&M3/M4\\
155.722837&61.321216&J102253.37+611917.5&MISDR2-00555-0771-0001&2&18.72&2.23&18.94&17.44&15.98&15.16&M3/M4\\   162.049490&64.724359&J104811.85+644328.2&MISDR1-00457-0489-0001&1&16.28&4.56&17.52&16.02&14.48&13.65&M3/M4\\
177.413955&0.586372&J114939.38+003510.2&MISDR1-13062-0283-0003&4&18.03&2.38&16.23&14.70&13.43&12.62&M2/M3 binary\\
179.486245&-0.667938&J115756.68-004004.6&MISDR1-13210-0285-0004&3&15.64&4.55&15.93&14.38&12.93&12.05&M3/M4\\
187.124063&12.375495& J122829.79+122231.9&NGA-Virgo-MOS02-0003&175&16.99&3.01&14.72&14.09&13.80&13.61&mid-K\\
188.802532&62.295738&J123512.63+621745.1&HDFN-00-0055&103&16.03&4.94&19.59&18.07&16.70&15.97&M3/M4\\
190.019963&62.385769&J124004.82+622309.1&HDFN-00-0019&103&18.61&2.11&22.40&21.08&19.29&18.41&M4/M5\\
199.682309&66.320356&J131843.62+661914&MISDR1-00384-0496-0001&11&18.11&2.83&23.01&21.57&20.10&19.23&M3/M4\\
201.406785&-2.761008&J132537.62-024540.9&MISDR1-33982-0341-0001&3&16.86&3.47&20.77&19.19&17.44&16.49&M4/M5\\
214.942778&52.995133&J141946.35+525942.5&GROTH-00-0062&169&17.49&3.13&15.64&14.18&13.17**&12.62&early M\\
215.043156&52.366049&J142010.42+522157.8&GROTH-00-0085&169&19.83&0.75&19.22&17.64&16.09&15.22&M4/M5\\
215.871036&4.171640&J142329.1+041017.6&MISDR1-33684-0584-0002&9&18.59&1.58&15.29&13.86&12.65**&11.95&early M\\
216.463311&4.496772&J142551.18+042948.7&MISDR1-33684-0584-0002&9&15.12&5.34&18.88&17.32&15.67&14.78&M4/M5\\
221.910686&3.886622&J144738.47+035312.1&MISDR1-33706-0587-0002&3&14.80&4.73&18.02&16.46&14.53&13.47&M5/M6\\
222.064056&3.373187&J144815.37+032223.3&MISDR1-33706-0587-0001&5&18.11&2.53&22.39&20.60&18.14&16.56&M7/M8\\
222.792793&31.111191&J145110.28+310639.5&UVE-A1979-0018&25&17.16&2.00&14.96**&11.96**&15.81**&10.26**&G 166-49, early M (vis. binary)***\\
222.792849&31.110976&J145110.28+310639.5&UVE-A1979-0001&25&17.34&1.88&14.96**&11.96**&15.81**&10.26**&G 166-49, early M (vis. binary)***\\
243.076275&52.706781&J161218.34+524225.2&NGA-NGC6090-0001&21&16.94&2.00&13.84&13.02**&12.64**&12.40&M star\\
251.924788&34.958986&J164741.89+345732.9&DEEPZLE-01-0014&37&18.10&2.19&16.18&14.74&13.48&12.81&M2/M3\\
252.243813&34.810881&J164858.54+344839.4&DEEPZLE-01-0015&37&17.56&2.46&14.66&16.50**&13.75&13.61&late K\\
253.162927&34.963021&J165239.24+345745.3&DEEPZLE-00-002846&46&17.60&2.30&16.72&15.33&15.06**&13.08&early M\\
258.318077&60.189264&J171316.24+601121.1&SIRTFFL-00-0013&40&18.25&2.34&23.31&21.80&20.00&18.92&M4/M5\\
259.443660&59.689733&J171746.57+594124.1&SIRTFFL-00-0019&30&17.33&2.93&14.90&13.49**&15.57**&12.19&M star\\
259.444028&59.690115&J171746.57+594124.1&SIRTFFL-00-0036&30&17.63&2.30&14.90&13.49**&15.57**&12.19&M star\\
260.306717&58.102743&J172113.52+580609.8&SIRTFFL-02-0004&11&18.69&1.99&20.23&18.68&17.34&16.57&M3/M4\\
309.706935&-1.194356&J203849.65-011140&MISDR2-19778-0981-0003&7&19.76&0.28&16.31&15.79&15.64&15.56&late G/early K\\
311.632980&0.414469&J204631.88+002452.5&MISDR2-19815-0981-0004&7&16.71&3.59&21.15&19.61&17.99&17.15&M4/M5\\
319.700768&0.314611&J211848.06+001849.7&MISDR2-20155-0986-0002&4&14.95&5.21&15.44&13.95&12.60**&14.09**&early M binary \\
330.570836&-7.378780&J220217.08-072243.3&MISDR2-20975-0716-0001&3&18.50&2.26&17.78&16.33&15.59&15.19&M0/M1\\
338.344710&-9.615278&J223322.67-093656.1&MISDR2-29639-0721-0005&6&16.91&1.81&16.67&24.80**&16.75**&22.83**&GJ 4282 sp=M3\\
340.440195&-1.094330&J224145.63-010539.2&MISDR2-21286-0377-0005&6&18.91&1.60&22.40&20.74&18.77&17.67&M5/M6\\
343.068027&-8.711330&J225216.32-084240.5&MISDR2-29544-0723-0001&3&15.81&4.63&17.65&16.15&14.71&13.95&M3/M4\\
345.574756&13.463785&J230217.85+132750&MISDR2-21073-0742-0001&1&18.85&1.60&14.44&13.77**&13.07**&12.50&late K/early M\\
347.307917&0.314596&J230913.89+001852.2&MISDR1-29084-0381-0002&6&18.35&2.33&21.78&20.46&18.74&17.82&M4/M5\\
350.754721&-0.021897&J232301.12-000118&MISDR1-29113-0383-0001&3&17.63&2.96&21.37&19.73&18.29&17.53&M3/M4\\
358.323456&15.050517&J235317.65+150302.1&MISDR2-28659-0749-0001&4&15.55&2.13&11.22**&10.66**&10.44**&11.81**&GSC 01722-00862, K-star\\
\hline
\hline
&&&&&&&&&&\\
\multicolumn{12}{l}{*cr = SDSS   image cosmic-ray contaminated, **= SDSS   photometric magnitude uncertain due to CCD saturation, *** = spectral type based on USNOB data} \\
\hline
\hline
\end{tabular}
\end{flushleft}
\end{tiny}
\end{table*}
}
\end{landscape}

\clearpage
}

\begin{landscape}
{
\renewcommand{\arraystretch}{1.1}
\begin{table*}[htbp]
\caption{\label{obslog} M-dwarf distances, NUV luminosities and flare energies }
\begin{tiny}
\begin{flushleft}
\begin{tabular}{lccccc}
\hline
\hline
SDSS   I.D.& Spectral Type& distance (pc)&log L$_{bol}$/L$_{\sun}$&E$_{NUVflare}$&log L$_{NUVflare}$/L$_{bol}$\\ 
\hline
J023955.52-072855.4&M0/M1&990&-1.27&     9.3e+30 &     -2.6 \\
J034358.01-053526.5&M3/M4&230&-1.99&     4.3e+29 &     -4.0 \\
J085939.62+565459.5&M1/M2&240&-1.54&     1.9e+30 &     -3.3 \\
J090350.46+522125&M5/M6&190&-2.81&       2.3e+29 &     -4.2 \\
J094659.05+613146.2&M5/M6&130&-2.81&     2.7e+29 &     -4.2 \\
J095758.82+020503.5&M3/M4&120&-1.99&     1.0e+29 &     -4.6 \\
J100141.6+020758.7&M2/M3&105&-1.72&      1.8e+29** &     -4.3** \\
J100141.6+020758.7&M2/M3&105&-1.72&      1.6e+29** &     -4.4 **\\
J100209.55+020726.1&M3/M4&405&-1.99&     6.1e+30 &     -2.8 \\
J100255.43+021159.6&M2/M3&790&-1.72&     3.2e+30 &     -3.1 \\
J101152.3+614454.1&M3/M4&70&-1.99&       6.0e+29 &     -3.8 \\
J101326.83+342309.8&M3/M4&505&-1.99&     4.1e+30 &     -3.0 \\
J102253.37+611917.5&M3/M4&130&-1.99&     2.8e+29 &     -4.1 \\
J104811.85+644328.2&M3/M4&60&-1.99&      1.8e+29 &     -4.3 \\
J114939.38+003510.2&M2/M3&40&-1.72&      2.3e+28 &     -5.2 \\
J115756.68-004004.6&M3/M4&25&-1.99&      1.2e+28 &     -5.5 \\
J123512.63+621745.1&M3/M4&260&-1.99&     1.1e+31 &     -2.5 \\
J124004.82+622309.1&M4/M5&460&-2.41&     5.5e+30 &     -2.8 \\
J131843.62+661914&M3/M4&690&-1.99&       1.6e+31 &     -2.4 \\
J132537.62-024540.9&M4/M5&145&-2.41&     3.2e+29 &     -4.1 \\
J142010.42+522157.8&M4/M5&110&-2.41&     4.5e+28 &     -4.9 \\
J142551.18+042948.7&M4/M5&85&-2.41&      7.6e+29 &     -3.7 \\
J144738.47+035312.1&M5/M6&25&-2.81&      9.2e+28 &     -4.6 \\
J144815.37+032223.3&M7/M8&40&-3.40&      8.1e+27 &     -5.7 \\
J164741.89+345732.9&M2/M3&75&-1.72&      1.9e+29 &     -4.3 \\
J171316.24+601121.1&M4/M5&280&-2.41&     1.1e+30 &     -3.6 \\
J172113.52+580609.8&M3/M4&295&-1.99&     1.3e+30 &     -3.5 \\
J204631.88+002452.5&M4/M5&295&-2.41&     7.3e+30 &     -2.7 \\
J220217.08-072243.3&M0/M1&600&-1.27&     5.4e+30 &     -2.8 \\
J224145.63-010539.2&M5/M6&145&-2.81&     1.8e+29 &     -4.3 \\
J225216.32-084240.5&M3/M4&90&-1.99&      1.2e+30 &     -3.5 \\
J230913.89+001852.2&M4/M5&300&-2.41&     1.0e+30 &     -3.6 \\
J232301.12-000118&M3/M4&475&-1.99&       4.4e+30 &     -2.9 \\
\hline
\hline
&&&&&\\
\multicolumn{6}{l}{** = 2 flares observed on this star} \\
\hline
\hline
\end{tabular}
\end{flushleft}
\end{tiny}
\end{table*}
}
\end{landscape}
\clearpage

\end{document}